\global\def\draftcontrol{0}

   \def\versionno{exotic thermodynamics }

\catcode`\@=11

\expandafter\ifx\csname draftcontrol\endcsname\relax\global\def\draftcontrol{0}
\fi

{\count255=\time\divide\count255 by 60
\xdef\hourmin{\number\count255}
\multiply\count255 by-60\advance\count255 by\time
\xdef\hourmin{\hourmin:\ifnum\count255<10 0\fi\the\count255}}
\def\draftdate{\number\month/\number\day/\number\year\ \ \ \hourmin }

\newcommand\makepapertitle{\par
  \begingroup
    \renewcommand\thefootnote{\@fnsymbol\c@footnote}%
    \def\@makefnmark{\rlap{\@textsuperscript{\normalfont\@thefnmark}}}%
    \long\def\@makefntext##1{\parindent 1em\noindent
            \hb@xt@1.8em{%
                \hss\@textsuperscript{\normalfont\@thefnmark}}##1}%
     \newpage
     \global\@topnum\z@   
     \@makepapertitle
     \thispagestyle{empty}\@thanks
  \endgroup
  \setcounter{footnote}{0}%
  \global\let\thanks\relax
  \global\let\makepapertitle\relax
  \global\let\@makepapertitle\relax
  \global\let\@thanks\@empty
  \global\let\@author\@empty
  \global\let\@date\@empty
  \global\let\@title\@empty
  \global\let\title\relax
  \global\let\author\relax
  \global\let\date\relax
  \global\let\and\relax
  \def\version{\let\version\@version\@gobble}
}
\def\@makepapertitle{%
  \newpage
   \ifnum\draftcontrol=1 {}
   \version\versionno
   \vskip 3em%
   \else
   \hfill\hbox to 3cm {\parbox{4cm}{\@pubnum}\hss}%
   \vskip 3em%
   \fi
   \begin{center}%
   \let \footnote \thanks
     {\LARGE {\@title}}%
     \vskip 1.5em%
     {\normalsize
       \lineskip .5em%
       \begin{tabular}[t]{c}%
         \@author
       \end{tabular}\par}%
     \vskip 1.5em%
     {\@bstract}%
     \end{center}%
     \vskip 1.5em
     \@date%
   \par
}

\gdef\@pubnum{}
\def\pubnum#1{%
  \gdef\@pubnum{#1}}

\gdef\@bstract{}
\def\Abstract#1{%
  \gdef\@bstract{%
   \parbox{\textwidth-0pc}{%
   \centerline{\bf Abstract}\penalty1000%
\kern.2cm%
\noindent
\renewcommand\baselinestretch{1.0}%
{#1}}}
}

\def\ps@paper{\let\@mkboth\@gobbletwo%
     \ifnum\draftcontrol=1
    \def\@oddfoot{\hbox to \textwidth{\tiny \versionno \hfil\tiny\draftdate}%
    \hskip -\textwidth \hbox to \textwidth{\hfil\rm\thepage\hfil}}%
     \else\def\@oddfoot{\hbox to \textwidth{\hfil\rm\thepage\hfil}}
     \fi
     \let\@evenfoot\@oddfoot
}

\def\body{\clearpage
          \pagestyle{paper}
    }

\def\@version#1{\ifnum\draftcontrol=1
\typeout{}\typeout{#1}\typeout{}
\vskip3mm\centerline{\hbox{\fbox{\normalsize{\tt DRAFT -- #1 -- }
                   {\draftdate}}}}\vskip3mm
\fi}
\let\version\@version
\long\def\eqlabel#1{\ifnum\draftcontrol=1
                    \tag@false  
                    \tag*{(\theequation) \hbox to -0.2cm{\hspace{0cm}\small{#1}\hss}}
                    \refstepcounter{equation}
                    \edef\@currentlabel{\theequation}
                    \ltx@label{#1}          
                    \else
                    \label{#1}
                    \fi
                    }
\let\st@bibitem\@bibitem
\let\st@lbibitem\@lbibitem
\ifnum\draftcontrol=1
  \def\@bibitem#1{%
    \st@bibitem{#1}\a@@label{#1}\ignorespaces}
  \def\@lbibitem[#1]#2{%
    \st@lbibitem[#1]{#2}\a@@label{#2}\ignorespaces}
  \def\a@@label#1{%
    \gdef\a@lab{\smash{\normalfont\small#1}}
    \ifvmode
      \if@inlabel
        \global\setbox\@labels\hbox{%
          \llap{\a@lab\let\a@lab\relax
                \kern\@totalleftmargin\kern\marginparsep}%
          \box\@labels}%
      \fi
    \fi}
\fi

\documentclass[12pt,letterpaper]{article}

\usepackage{amsmath,amssymb,array,calc,epsfig}
\usepackage{graphicx}
\usepackage{psfrag,verbatim,bm}

\ifnum\draftcontrol=1
\fi

\renewcommand\baselinestretch{1.25}
\renewcommand\section{\@startsection {section}{1}{\z@}%
                                   {-3.5ex \@plus -1ex \@minus -.2ex}%
                                   {2.3ex \@plus.2ex}%
                                   {\normalfont\large\bfseries}}
\renewcommand\subsection{\@startsection{subsection}{2}{\z@}%
                                   {-3.25ex\@plus -1ex \@minus -.2ex}%
                                   {1.5ex \@plus .2ex}%
                                   {\normalfont\normalsize\bfseries}}
\renewcommand\subsubsection{\@startsection{subsubsection}{3}{\z@}%
                                   {-3.25ex\@plus -1ex \@minus -.2ex}%
                                   {1.5ex \@plus .2ex}%
                                   {\normalfont\normalsize\it}}
\renewcommand\paragraph{\@startsection{paragraph}{4}{\z@}%
                                   {-3.25ex\@plus -1ex \@minus -.2ex}%
                                   {1.5ex \@plus .2ex}%
                                   {\normalfont\normalsize\bf}}

\numberwithin{equation}{section}

\def\ie{{\it i.e.}}

\def\revise#1       {\raisebox{-0em}{\rule{3pt}{1em}}%
                     \marginpar{\raisebox{.5em}{\vrule width3pt\
                     \vrule width0pt height 0pt depth0.5em
                     \hbox to 0cm{\hspace{0cm}{%
                     \parbox[t]{4em}{\raggedright\footnotesize{#1}}}\hss}}}}

\newcommand\nxt[1]  {\\\fnxt#1}

\def\cale         {{\cal E}}
\def\calf         {{\cal F}}

\def\calh         {{\cal H}}

\def\call         {{\cal L}}

\def\calo         {{\cal O}}

\def\zet          {{\mathbb Z}}

\def\sqr#1#2{{\vcenter{\vbox{\hrule height.#2pt
 \hbox{\vrule width.#2pt height#1pt \kern#1pt
 \vrule width.#2pt}\hrule height.#2pt}}}}

\def\a{\alpha}

\def\l{\lambda}

\def\c{\chi}

\def\p{\phi}

\catcode`\@=12

\begin{document}


\title{\bf Exotic Hairy Black Holes}
\pubnum
{UWO-TH-09/7
}

\date{April 2009}

\author{
Alex Buchel$ ^{1,2}$ and Chris Pagnutti$ ^{1}$\\[0.4cm]
\it $ ^1$Department of Applied Mathematics\\
\it University of Western Ontario\\
\it London, Ontario N6A 5B7, Canada\\
\it $ ^2$Perimeter Institute for Theoretical Physics\\
\it Waterloo, Ontario N2J 2W9, Canada\\
}

\Abstract{
We study black hole solutions in asymptotically $AdS_4$ spacetime with
scalar hair.  Following AdS/CFT dictionary these black holes can be
interpreted as thermal states of 2+1 dimensional conformal gauge
theory plasma, deformed by a relevant operator.  We discover a rich
phase structure of the solutions. Surprisingly, we find
thermodynamically stable phases with spontaneously broken global
symmetries that exist only at high temperatures. These phases are
metastable, and join the stable symmetric phase via a mean-field
second-order phase transition.
}

\makepapertitle

\body

\version\versionno

\section{Introduction}
The power of holographic 
gauge theory/string theory correspondence of Maldacena \cite{m9711} is that it extends beyond the 
conformal examples, and thus can be used as a tool to study critical phenomena in strongly coupled 
systems. Furthermore, while there are few explicit realizations of this duality in a non-conformal setting, 
the application of the {\it effective} Maldacena framework provides a new valuable (phenomenological)
tool in addressing the strongly coupled dynamics. Such a framework was applied with intriguing results to 
formulate the holographic model of QCD \cite{g1}, non-relativistic AdS/CFT correspondence \cite{g2,g3},
effective description of viscous hydrodynamics \cite{g4,g5}, holographic theories of 
superfluidity \cite{g6} and superconductivity \cite{g7}, to name some.
Often a posteriori, these phenomenological models of the Maldacena correspondence were embedded in 
string theory.
The application of the holographic framework itself, rather than its specific string theory realization, 
is particularly relevant to the physics of the continuous critical phenomena where one expects
organization of the physical phenomena in universality classes, thus obliterating the importance of 
any particular representative in a universality class. While it is certainly interesting to 
find embedding of the models discussed here in string theory, the latter is not 
our primary objective in this paper.

In this paper we follow up on a very interesting suggestion by 
Gubser\footnote{We would like to thank Omid Saremi for bringing this work to our attention and for 
valuable discussions concerning his related work \cite{saremi}.} \cite{gubser} and will attempt to 
engineer finite temperature critical phenomena in non-conformal 2+1 dimensional strongly coupled 
systems\footnote{Our choice of the dimensionality was dictated by the further possibility to study 
these systems (in a holographic setting) in external magnetic fields \cite{kh}.}.  
We now proceed to discuss the motivation of our model from the field-theoretic perspective. 
Consider a 2+1 dimensional relativistic conformal field theory. A deformation of this CFT by a relevant 
operator $\calo_r$, 
\begin{equation}
\calh_{CFT}\ \to\ \tilde{\calh}=\calh_{CFT}+\lambda_r \calo_r
\eqlabel{deform}
\end{equation}
softly breaks the scale invariance and induces the renormalization group flow.  
Our experience with similar deformations \cite{n21,n22} suggests that in a thermal state 
 $\calo_r$ will develop a vacuum expectation value. This expectation value tends to  
zero in a high temperature phase, but it  grows as the temperature $T$ becomes comparable to a scale 
$\Lambda$,
set by the coupling $\lambda_r$. We assume that the deformed theory $\tilde{\calh}$ has an 
irrelevant operator $\calo_{i}$. Note that we do not (further) deform $\tilde{\calh}$ by 
$\calo_{i}$, as this would destroy a UV fixed point defined by $\calh_{CFT}$. We consider the case 
in which the operator $\calo_{i}$ has an associated  $\zet_2$ discrete symmetry. As long 
as the operators $\calo_r$ and $\calo_i$ do not mix under the RG flow, the expectation value 
of $\calo_i$ is zero. Of course, the generic situation is that the operator mixing is 
present, in which case $\calo_i$ will develop a vev in a thermal state. The absence of 
discrete symmetries would typically produce a nonvanishing expectation value of irrelevant 
operators even in the high temperature phase \cite{kt1,kt2}. On the other hand, it is natural 
to expect that a  discrete symmetry (preserved by a RG flow) 
would be protected at least when the operator mixing is small, \ie, in the high
temperature phase. 
As $T\lesssim  \Lambda$ one expects that the mixing would be large, producing the condensate 
of $\calo_i$ and thus spontaneously breaking $\zet_2$ symmetry. A holographic implementation 
of this phenomenon was proposed in \cite{gubser}.  
   
Rather surprisingly, we find that in our specific holographic realization of Gubser's 
proposal \cite{gubser}, spontaneous breaking of a discrete symmetry occurs at high temperature.
The thermal phases with the broken symmetry exist only above certain critical temperatures ---
different for each distinct broken phase. In a holographic setting, each broken phase is 
identified by the number of nodes in the wavefunction of the supergravity field dual to 
an irrelevant operator $\calo_i$.
Although thermodynamically stable, these phases have higher free energy than the symmetric
phase. Some of them connect to a symmetric phase via a second-order mean-field 
transition, with $\langle \calo_i\rangle$ as an order parameter.       

In the next section we discuss our holographic model in detail. The results of the 
thermodynamic analysis are presented in section 3. We conclude in section 4 with future directions 
and further conjectures.

\section{A holographic model of an exotic critical phenomena}

In this section we discuss an explicit realization of Gubser's holographic model of critical 
phenomena \cite{gubser}. 

The effective four-dimensional gravitational bulk action, dual to a field-theoretic setup 
discussed in the introduction, takes form\footnote{We set the radius of an asymptotic $AdS_4$ 
geometry to unity.} 
\begin{equation}
\begin{split}
S_4=&S_{CFT}+S_{r}+S_i=\frac{1}{2\kappa^2}\int dx^4\sqrt{-\gamma}\left[\call_{CFT}+\call_{r}+\call_i\right]\,,
\end{split}
\eqlabel{s4}
\end{equation}
\begin{equation}
\call_{CFT}=R+6\,,\qquad \call_r=-\frac 12 \left(\nabla\phi\right)^2+\phi^2\,,\qquad 
\call_i=-\frac 12 \left(\nabla\chi\right)^2-2\chi^2-g \phi^2 \chi^2 \,,
\eqlabel{lc}
\end{equation}
where we split the action into (a holographic dual to)  a CFT part $S_{CFT}$; its deformation by a relevant
operator $\calo_r$; and a sector $S_i$ involving an irrelevant operator $\calo_i$ 
along with its mixing with $\calo_r$ under the RG dynamics.
The four dimensional gravitational constant $\kappa$ is related to a central charge $c$ of the 
UV fixed point as 
\begin{equation}
c=\frac{192}{ \kappa^2}\,.
\eqlabel{cg}
\end{equation}
In our case the scaling dimension of $\calo_r$ is either 
1 or 2, depending on which of the two normalizable boundary modes of a gravitational scalar $\phi$ 
\cite{phi1}
we treat 
as a field-theoretic parameter (and thus keep it constant as we vary the temperature), 
and the scaling dimension of $\calo_i$ is 4.  In order to have asymptotically $AdS_4$ solutions, 
we assume that only the normalizable mode of $\calo_i$ is nonzero near the boundary.
Finally, we assume that $g<0$ in order to holographically induce the critical behavior 
\cite{gubser}.

Effective action \eqref{s4} has a $\zet_2\times \zet_2$ discrete symmetry that acts as a parity transformation on 
the scalar fields $\phi$ and $\chi$. The discrete symmetry $\phi\to -\phi$ is softly broken by a relevant deformation 
of the $AdS_4$ CFT; while as we will see, the $\chi\to -\chi$ symmetry is broken spontaneously.

We will study Schwarzschild black hole solutions in \eqref{s4} with translationary invariant 
horizons and nontrivial scalar hair. Thus, we assume:
\begin{equation}
ds_4^2=-c_1(r)^2\ dt^2+c_2(r)^2\ \left[dx_1^2+dx_2^2\right]+c_3^2\ dr^2\,,\qquad \phi=\phi(r)\,,\qquad 
\chi=\chi(r)\,.
\eqlabel{background}
\end{equation}
As in \cite{kt1}, for the precision numerical analysis we find it convenient to introduce
a new radial coordinate $x$ as follows
\begin{equation}
1-x\equiv \frac{c_1(r)}{c_2(r)}\,,
\eqlabel{xdef}
\end{equation} 
so that $x\to 0$ corresponds to the boundary asymptotic, and $y\equiv 1-x\to 0$ 
corresponds to a regular Schwarzschild horizon asymptotic.
Further introducing 
\begin{equation}
c_2(x)=\frac{a(x)}{(2x-x^2)^{1/3}}\,,
\eqlabel{defa}
\end{equation}
the equations of motion obtained from \eqref{s4}, \eqref{lc} with the background 
ansatz \eqref{background}, and the  boundary conditions discussed above, imply 
\begin{equation}
\begin{split}
a=&\a\left(1-\frac{1}{40}\ p_1^2\ x^{2/3}-\frac{1}{18}\ p_1 p_2\ x+\calo(x^{4/3})\right)\,,\\
\p=&p_1\ x^{1/3}+p_2\ x^{2/3}+\frac{3}{20} p_1^3 x+\calo(x^{4/3})\,,\\
\c=&\c_4\left(x^{4/3}+\left(\frac 17 g -\frac{3}{70}\right)p_1^2\ x^2+\calo(x^{7/3})\right)\,,
\end{split}
\eqlabel{boundary}
\end{equation}
near the boundary $x\to 0_+$, and 
\begin{equation}
\begin{split}
a=\a\left(a_0^{h}+a_1^h\ y^2+\calo(y^4)\right)\,,\qquad \p=p_0^h+\calo(y^2)\,,\qquad  \p=c_0^h+\calo(y^2)\,,
\end{split}
\eqlabel{hor}
\end{equation}
near the horizon $y=1-x\to 0_+$. 
Apart from the overall scaling factor $\a$ (which is related to the temperature), the background is 
uniquely specified with 3 UV coefficients $\{p_1,p_2,\c_4\}$ and 4 IR coefficients 
$\{a_0^h,a_1^h,p_0^h,c_0^h\}$. 
Recall that both modes of $\phi$ near the boundary are normalizable. Thus we have a freedom to interpret the 
relevant deformation \eqref{deform} as due to $\calo_r$ with $\dim[\calo_r]=2$, in which case we have to keep
the combination $p_1 \a\equiv \Lambda$ fixed, or due to $\calo_r$ with $\dim[\calo_r]=1$, in which case we have to keep 
$p_2 \a\equiv \Lambda^2$ 
fixed. We can determine these scalings by carefully matching the boundary asymptotic finite temperature background 
\eqref{background} to the 
boundary asymptotic of the zero temperature background  --- in the latter case it is clear what is meant to keep $\l_r$ 
fixed\footnote{Identical procedure has been used in \cite{kt1}.}. As we point out below, the highly nontrivial check 
on our identification is the consistency of the hairy black hole thermodynamics with the first law of thermodynamics.  
The scale $\Lambda$ is the scale set by the relevant coupling $\lambda_r$ in \eqref{deform}.
Notice that the two solutions 
\[
(p_1,p_2,\c_4,a_0^h,a_1^h,p_0^h,c_0^h)\qquad \&\qquad (-p_1,-p_2,-\c_4,a_0^h,a_1^h,-p_0^h,-c_0^h)
\]
are equivalent. Thus we can assume $p_1>0$ for $\dim[\calo_r]=2$, and $p_2>0$ for $\dim[\calo_r]=1$.

It is straightforward to compute the temperature $T$ and the entropy density $s$ of the black hole solution \eqref{background}
\begin{equation}
\left(\frac{8\pi T}{\a}\right)^2=\frac{6 (a^h_0)^3 (6-2 (\c^h_0)^2+(p^h_0)^2-g (p^h_0)^2 (c^h_0)^2)}{3 a^h_1+a^h_0}\,,
\eqlabel{tbh}
\end{equation}
\begin{equation}
s=\frac{c}{384}\ 4\pi\a^2\ (a_0^h)^2\,.
\eqlabel{sbh}
\end{equation}
The energy density $\cale$ and the free energy density $\calf$ can be computed using the 
techniques of the holographic renormalization 
\cite{holren}. There is a slight subtlety in holographic renormalization in our case associated with the fact that both modes 
of $\calo_r$ are normalizable near the boundary \cite{holren2}:
\nxt when $\dim[\calo_r]=2$ we have
\begin{equation}
\begin{split}
\hat{\calf}\equiv\frac{384}{c}\ \calf=&\a^3 \left(2-\frac 16  p_1 p_2-\frac {(a_0^h)^3}{2}
 \sqrt{\frac{6 a^h_0 (6-2 (c^h_0)^2+(p^h_0)^2-g (p^h_0)^2 (c^h_0)^2)}{3 a^h_1+a^h_0}}\right)\,,\\
\hat{\cale}\equiv\frac{384}{c}\ \cale=&\a^3 \left(2-\frac 16  p_1 p_2\right)\,,
\end{split}
\eqlabel{p1fixed}
\end{equation} 
\nxt when $\dim[\calo_r]=1$ we have 
\begin{equation}
\begin{split}
\hat{\calf}\equiv\frac{384}{c}\ \calf=&\a^3 \left(2+\frac 13  p_1 p_2-\frac {(a_0^h)^3}{2}
 \sqrt{\frac{6 a^h_0 (6-2 (c^h_0)^2+(p^h_0)^2-g (p^h_0)^2 (c^h_0)^2)}{3 a^h_1+a^h_0}}\right)\,,\\
\hat{\cale}\equiv\frac{384}{c}\ \cale=&\a^3 \left(2+\frac 13  p_1 p_2\right)\,.
\end{split}
\eqlabel{p2fixed}
\end{equation} 
In both cases we independently evaluated $\{\calf,\cale,s\}$. A nontrivial check on the holographic renormalization 
is the automatic fulfillment of the basic thermodynamic relation
\[
\calf=\cale - T s \,.
\]

The UV fixed point is described by $AdS_4$ Schwarzschild-black hole solution 
\begin{equation}
a(x)=\frac{\a}{(2x-x^2)^{1/3}}\,,\qquad \phi(x)=\chi(x)=0\,,
\eqlabel{fixed}
\end{equation}
in which case we have
\begin{equation}
T=\frac{3\a}{4\pi}\,,\qquad \frac{\hat{\calf}}{(\pi T)^3}=-\frac{64}{27}\,,
\qquad \frac{\hat{\cale}}{(\pi T)^3}=\frac{128}{27}\,.
\eqlabel{fixedth}
\end{equation}

We conclude this section with comments on the numerical procedure. We use numerical technique identical to the one developed 
in \cite{kt1}. For  $\dim[\calo_r]=2$ we vary $p_1$;  for each given value of $p_1$ we compute the remaining UV and the IR 
coefficients   $(p_2,\c_4,a_0^h,a_1^h,p_0^h,c_0^h)$. Notice that there are just the right number of these coefficients to 
uniquely determine the solution of a system of 3 second order differential equations (the equations of motion)
for the background functions $\{a(x)\,, \phi(x)\,, \chi(x)\}$. Given the numerical data, we can use \eqref{tbh} 
and the condition $p_1 \a =\Lambda$ to present the data, say for the free energy,  
as $\frac{\hat{\calf}}{T^3}$ versus $\frac \Lambda  T$. 
The procedure is then repeated for $\dim[\calo_r]=2$ with appropriate replacements of $p_1$ with $p_2$.

\section{Thermodynamics of hairy black holes in asymptotic $AdS_4$ geometry}

\begin{figure}[t]
\begin{center}
\psfrag{mu}{{$\frac{\Lambda}{8\pi T}$}}
\psfrag{dim2}{{$\dim[\calo_r]=2$}}
\psfrag{dim1}{{$\dim[\calo_r]=1$}}
\psfrag{f}{{$\frac{\hat{\calf}}{(\pi T)^3}$}}
  \includegraphics[width=3in]{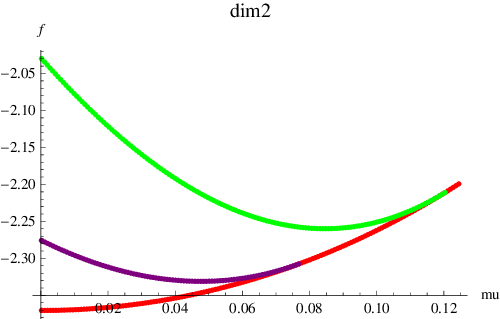}
\includegraphics[width=3in]{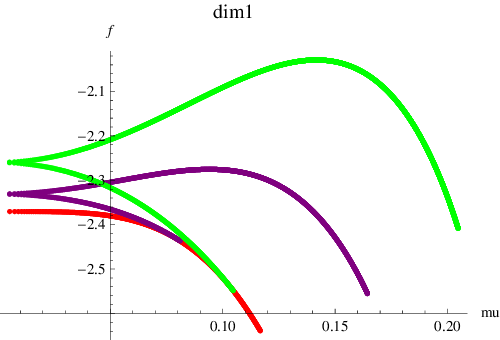}
\end{center}
  \caption{
The free energy density of hairy black holes in asymptotic $AdS_4$ geometry for different deformations of the 
UV fixed point. The red curves correspond to  symmetric phases with $\langle \calo_i\rangle =0$. The other curves 
correspond to phases with spontaneously broken $\zet_2$ symmetry, $\langle \calo_i\rangle \ne 0$. The purple 
curves represent symmetry broken phases with the holographic condensate wavefunction $\chi(x)$ not 
having any nodes for $x\in [0,1]$. The green
curves represent symmetry broken phases with the holographic condensate wavefunction $\chi(x)$ 
having exactly one node for $x\in [0,1]$. } \label{figure1}
\end{figure}

\begin{figure}[t]
\begin{center}
\psfrag{mu}{{$\frac{\Lambda}{8\pi T}$}}
\psfrag{dim2}{{$\dim[\calo_r]=2$}}
\psfrag{dim1}{{$\dim[\calo_r]=1$}}
\psfrag{f}{{$\frac{\hat{\cale}}{(\pi T)^3}$}}
  \includegraphics[width=3in]{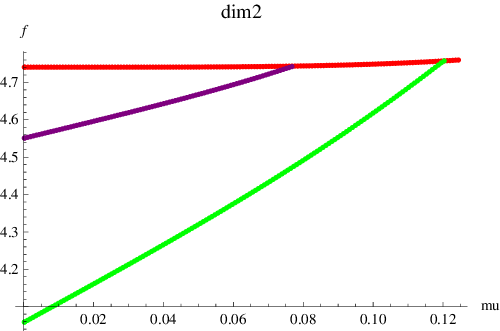}
\includegraphics[width=3in]{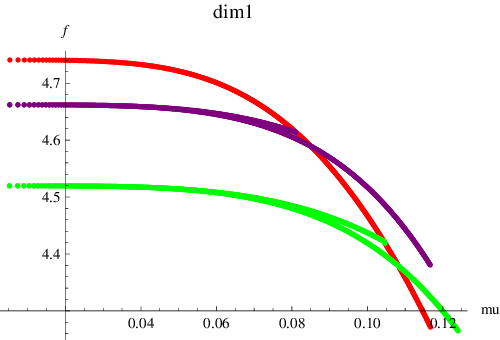}
\end{center}
  \caption{
The energy density of hairy black holes in asymptotic $AdS_4$ geometry for different deformations of the 
UV fixed point. The color coding is as in figure~\ref{figure1}. } \label{figure2}
\end{figure}

\begin{figure}[t]
\begin{center}
\psfrag{mu}{{$\frac{\Lambda}{8\pi T}$}}
\psfrag{dim2}{{$\dim[\calo_r]=2$}}
\psfrag{dim1}{{$\dim[\calo_r]=1$}}
\psfrag{v2}{{$c_s^2$}}
  \includegraphics[width=3in]{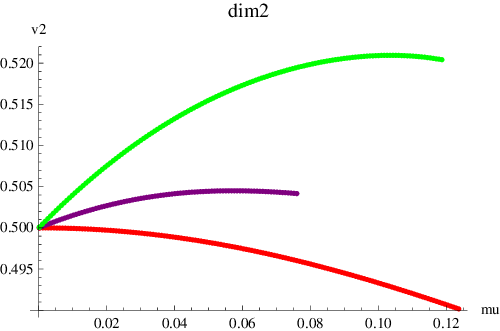}
\includegraphics[width=3in]{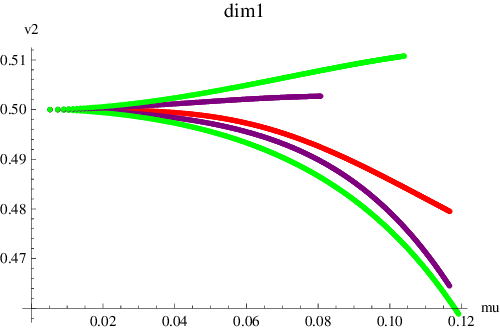}
\end{center}
  \caption{
The speed  of sound of hairy black holes in asymptotic $AdS_4$ geometry for different deformations of the 
UV fixed point. The color coding is as in figure~\ref{figure1}. } \label{figure3}
\end{figure}

\begin{figure}[t]
\begin{center}
\psfrag{lng}{{$\ln(-g)$}}
\psfrag{lnf}{\hspace{-1.8cm}{$\Delta=\ln\left(\frac{\hat{\calf}}{(\pi T)^3}+\frac{64}{27}\right)$}}
\psfrag{dim2}{{$\dim[\calo_r]=2$}}
  \includegraphics[width=3in]{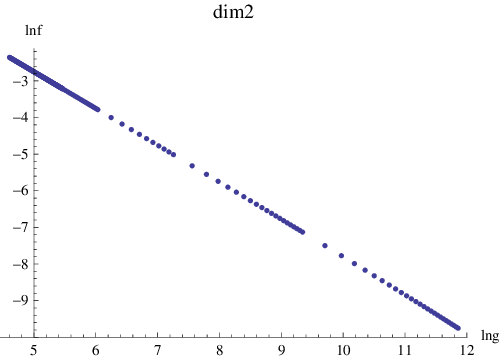}
\end{center}
  \caption{
The difference between the free energy density of the "purple'' symmetry broken phase 
and  the symmetric phase (see figure~\ref{figure1}) for $\frac\Lambda T\to 0$ as a function of $\ln(-g)$.
The straight line fit to the data produces the following functional dependence: $\Delta\approx 2.32861 - 1.01493 \ln(-g)$.
}\label{figure4}
\end{figure}

In this section we present results of the thermodynamic analysis of the hairy black holes.
We choose $g=-100$ in \eqref{lc}.

Figures.~\ref{figure1} and  \ref{figure2} represent the free energy density and the energy density (correspondingly)
of the hairy black holes in asymptotic $AdS_4$ geometry. Each plot is labeled by the dimension of the relevant operator 
$\calo_r$ used to generate the holographic RG flow, see \eqref{deform}. The red curves correspond to  $\zet_2$ 
symmetric phases with a vanishing expectation value of the irrelevant operator $\calo_i$ (recall in our model 
 $\dim[\calo_i]=4$ ). The remaining curves represent phases with the broken   $\zet_2$  symmetry, and 
the non-vanishing expectation value for $\calo_i$. The purple curves denote symmetry broken phases in which the 
wavefunction of the gravitational holographic dual to $\calo_i$ --- $\chi(x)$ --- has no nodes for  $x\in [0,1]$. 
 The green curves denote symmetry broken phases in which the 
wavefunction $\chi(x)$ has exactly one node for  $x\in [0,1]$.

Some of the comments below follow from the data presented; some other comments rely on the results of the 
analysis we omit here. 
\nxt We verified the consistency of our analysis and the identification of the mass scale in deformation 
\eqref{deform} by checking the first law of thermodynamics:
\[
d\calf=-s dT\,.
\]
In our numerical results
\begin{equation}
\bigg|1+\frac{d\calf}{s d T}\bigg|\lesssim 10^{-7}\,.
\eqlabel{error}
\end{equation}
\nxt All the symmetry broken phases for the deformation with $\dim[\calo_r]=2$ 
(and some of the phases  generated by the  $\dim[\calo_r]=2$ RG flow) exist only above certain 
critical temperature, $T_c$. Notice that this $T_c$ is different for phases in which the condensate 
$\chi(x)$ has different number of nodes ---  $T_c^{purple}<T_c^{green}$  (recall that in  purple phases 
$\chi(x)$ does not have any nodes, and in green phases $\chi(x)$ has one node ).  
\nxt All the symmetry broken phases have a higher free energy density compare to a symmetric phase. 
Thus, at best, they are metastable. Further work is needed to determine if phases with spontaneous symmetry breaking 
are stable \cite{wip}. Note that if these phases are shown to be perturbatively unstable
(having a tachyon in the spectrum of fluctuations), it will be rather unusual given that these 
phases survive to arbitrary high temperatures $\frac{\Lambda}{T}\to 0$. 
\nxt All the broken phases for the  $\dim[\calo_r]=2$ deformation, and the phases for the $\dim[\calo_r]=1$ 
deformation which ``end'' on the symmetric phase (the red curves), undergo a mean-field second-order
phase transition. In all cases as $T\to T_c^+$ 
\begin{equation}
\langle \calo_i\rangle \propto \c_4\propto \left(T-T_c\right)^{1/2}\,,
\eqlabel{transition}
\end{equation}
\begin{equation}
\hat{\calf}^{broken}-
\hat{\calf}^{symmetric}\propto\begin{cases}  +(T-T_c)^2\,, & \text{as $T>T_c$\,,}
\\
0 &\text{as $T\le T_c$\,.}
\end{cases}
\eqlabel{deltaf}
\end{equation}
\nxt All the phases we discuss are thermodynamically stable, at least in the temperature 
range under consideration. Figure~\ref{figure3} 
represents the dependence of the 
speed of sound squared $c_s^2$ on $\frac{\Lambda}{T}$.
\nxt We found symmetry broken phases with  $\chi(x)$ having more than one mode for $x\in [0,1]$, thus we believe 
there are infinitely many high temperature phases with $\langle \calo_i\rangle \ne 0$. We expect that all these phases 
share the same general features as the phases we discussed in details above. 
\nxt The data in figures~\ref{figure1}-\ref{figure3} correspond to RG flows in which the mixing between 
$\calo_r$ and $\calo_i$ (the coupling constant $g$ in \eqref{lc}) is fixed, and negative. 
We studied the dependence of the thermodynamics on $g$.
As $|g|$ decreases, the broken phase free energy density evaluated at  a given temperature increases. This increase is particularly dramatic
as $g\sim -4\cdots -3$ (for a purple phase) 
suggesting that there is a critical 
mixing $g_c$ (presumably different for different broken phases) at which this
 symmetry breaking thermodynamic 
phase disappears. We did not find any symmetry broken phases for positive $g$,
thus it is likely that $g_c<0$. Perhaps even more interesting is the behavior of the symmetry broken 
phases for large mixing between $\calo_r$ and $\calo_i$, \ie, when $g\to -\infty$.  In 
figure~\ref{figure4} we show that as the mixing between the operators along the RG flow increases, the broken 
phases approach the symmetric phase at high temperatures. The latter fact is evident both in the behavior of the 
free energy density and in the vanishing of $\langle \calo_i\rangle$ for the high temperature broken phases as $g\to -\infty$.

\section{Conclusion}

In this paper we used the ideas of Gubser \cite{gubser} to holographically engineer RG flows in strongly coupled $2+1$ 
dimensional gauge theory models with exotic finite temperature phases. Specifically, we considered a relevant deformation of 
a relativistic $CFT_3$ with an operator $\calo_r$, $\dim[\calo_r]=\{1,2\}$. We allowed a mixing of an irrelevant operator 
$\calo_i$, $\dim[\calo_i]=4$, with $\calo_r$ along the RG flow. We discovered a plethora of exotic phases in the system at finite 
temperature. We found the phases that exist only at high temperature, spontaneously break $\zet_2$ symmetry of the model
and join the symmetric phase with a mean-field second-order phase transition. All the phases we discuss are thermodynamically stable.
The phases with the broken symmetry have a higher free energy density than the symmetric phase; thus, at best, they are metastable.

Still a lot of work (both technical and conceptual) is needed to understand these exotic phases.
\nxt  It would be interesting to study perturbative stability of the phases with the broken symmetry. 
Using gauge theory/ string theory correspondence it is easy to argue that thermodynamic instabilities of the 
gravitational backgrounds reveal themselves as  (perturbative) dynamical instabilities \cite{bgm} as well.
The reverse statement is not necessarily true.
For such analysis one needs to compute the spectrum of the quasinormal modes associated with the 
fluctuations of $\chi(x)$. 
\nxt We made a lot of conjectures about the phase structure of the theory based on extensive numerical work.
It would be interesting to analytically prove the disappearance of the symmetry breaking phases 
for small mixing, $g>g_c$. Likewise, it is important to show analytically that there is no 
phase crossing in the system as $g\to -\infty$. If such a phase crossing does occur, the high temperature 
symmetry broken phases would be thermodynamically preferable; in addition, there could be interesting 
critical behavior associated with such crossing.   
\nxt We did not study the question of embedding the RG flows discussed here in string theory ---
it might very well be that there is an (unknown to us at this stage) obstruction for such an embedding.
It would be interesting to see if similar phenomena occur in string theory realizations 
of holographic dualities of seemingly related systems \cite{pw2}. 
\nxt We want to emphasize once again the unusual features of the phases with the broken symmetry.
It is well known that a two-dimensional Ising model allows for a dual reformulation in terms of 
a "disorder'' operator\footnote{See \cite{kogut} for a review.}. The Hamiltonian of the 
Ising model is local when expressed in terms of the disorder operators.  While the order parameter in
the spin formulation vanishes above certain temperature, in the disorder formulation, the disorder 
parameter (probing the condensation of the kinks) 
vanishes in the low temperature phase, akin to what we see in our model here.  
The relation between 
the original  spin and the dual disorder operator in the Ising model is nonlocal. In our model 
both the operators $\calo_r$ and $\calo_i$ are {\it local}. If $\calo_i$ is interpreted as an analog of the 
 disorder operator in the Ising model, how can it be mutually local with $\calo_r$ ---  an analog of a local 
operator in the spin formulation of the Ising model?  
\nxt  There are known condensed matter systems which have a disordered high temperature phase 
\cite{cond}. It would be interesting to study if a close analogy between those systems and the 
holographic models discussed here could be established.
\nxt Finally, it would be interesting to develop in more details the idea of  'microengineering' the 
renormalization group flow by specifying not only the operators that deform the (UV) fixed point
of the theory, or simply are present in the theory, but also imposing --- by hand --- the operator mixing under the renormalization group flow.
While this is quite unusual in a standard field-theoretic setting, it is a real possibility in the 
holographic constructions.

\section*{Acknowledgments}
We would like to thank Zhenya Buchbinder, Colin Denniston, Jaume Gomis, Volodya Miransky 
and Omid Saremi for valuable discussions. 
Research at Perimeter Institute is
supported by the Government of Canada through Industry Canada and by
the Province of Ontario through the Ministry of Research \&
Innovation. AB gratefully acknowledges further support by an NSERC
Discovery grant and support through the Early Researcher Award
program by the Province of Ontario. CP acknowledges support by NSERC.

\end{document}